\documentclass[aps,prb,twocolumn,amsmath,amssymb]{revtex4}

\usepackage{graphicx}
\usepackage{bbold}
\usepackage[utf8]{inputenc}

\begin{document}
\title{Electronic heat current fluctuations in a quantum dot}
\author{A.~Cr\'epieux$^{1}$}
\affiliation{$^1$ Aix Marseille Univ, Universit\'e de Toulon, CNRS, CPT, Marseille, France}

\parindent = 0pt

\begin{abstract}
The fluctuations of the heat current in a quantum dot coupled to electron reservoirs are calculated at finite frequency, voltage and temperature using the nonequilibrium Green function technique. The non-symmetrized heat noise is expressed as an integral on energy containing four contributions, each of which includes transmission amplitudes, electron-hole pair distribution functions and energy difference factors. The effect of the asymmetry of the couplings between the quantum dot and the reservoirs is studied. Features of the heat noise are highlighted and discussed for an equilibrium and an out-of-equilibrium quantum dot. In the latter case and within the high transmission limit, the heat noise is closely related to the radiative power spectrum, leading to an out-of-equilibrium Planck's law. Proposals for the measurement of the heat noise are discussed.
\end{abstract}

\maketitle

\section{Introduction}

In quantum devices, the heat fluctuates over time for several reasons: the first one is related to the presence of thermal agitation at finite temperature, the second one to the fact that the device interacts with its electromagnetic environment by emitting or absorbing energy via phonons or photons,  and the third one to the probabilistic nature of particle transfer in quantum systems. The characterization of these heat fluctuations provides a variety of information on energy dissipation\cite{Yu2016}, the presence of finite coherence and entanglement in open quantum systems\cite{Silaev2014}, and the higher-order cumulants of charge counting statistics\cite{Kindermann2004}. In addition, they reveal features that are not visible in the charge noise\cite{Battista2014} such as the signature of a crossover from Coulomb blockade to Kondo physics in energy fluctuations\cite{Ridley2019}. In the case of on-demand single-electron sources, the heat fluctuates while the charge emission is noiseless\cite{Battista2013}. So far, only temperature fluctuations\cite{Heikkila2009}, related to energy fluctuations\cite{Battista2013,Berg2015,Dashti2018}, have been measured\cite{Karimi2020}, but there are several proposals for the measurement of heat fluctuations\cite{Laakso2012,Sanchez2012,Berg2015,Dashti2018}. With the fast progress of heat measurement techniques in nanosystems\cite{Meschke2006,Timofeev2009,Ciliberto2013,Sivre2019}, it can be expected that this will be possible in the foreseeable future. Heat transport in quantum devices is in itself well controlled\cite{Jezouin2013}, notably with the experimental confirmation\cite{Schwab2000} of the existence of a thermal conductance quantum\cite{Pendry1983}, and the evidence of the heat Coulomb blockade effect\cite{Sivre2018}.

The issues raised by these studies are also of fundamental interest. The question of the generalization of the fluctuation-dissipation theorem to heat transport has been addressed\cite{Averin2010,Crepieux2017,Pekola2018}, as well as the verification of the fluctuation theorem\cite{Jarzynski2004,Esposito2009,Averin2011,Sanchez2012,Rahav2012,Utsumi2014}, which is a microscopic extension of the second law of thermodynamics. The statistics of heat exchange in a driven open quantum system have been studied\cite{Gasparinetti2014} as well as the statistics of work for a two-level system in the presence of dissipation\cite{Hekking2013}. Among the theoretical approaches used to study the heat fluctuations in quantum devices, one can cite the Landauer-B\"uttiker formalism\cite{Blencowe1999,Sergi2011}, nonequilibrium Schwinger-Keldysh Green function technique\cite{Kindermann2004,Clerk2011,Zhan2011,Zhan2013}, circuit theory\cite{Averin2011}, Tomonaga-Luttinger liquid theory\cite{Krive2001,Ronetti2019}, generalized Lindblad master equations\cite{Sanchez2012,Rahav2012,Silaev2014}, mean-field method coupled to Hartree-Fock approximation\cite{Wu2009}, Boltzmann-Langevin approach\cite{Dashti2018}, and inchworm quantum Monte Carlo method\cite{Ridley2019}. The systems in question are either molecular junctions\cite{Wu2009,Zhan2011}, quantum wires\cite{Blencowe1999,Krive2001,Agarwalla2012}, mesoscopic constrictions\cite{Averin2010}, quantum dots\cite{Averin2011,Sanchez2012,Sanchez2013,Zhan2013,Eymeoud2016,Yu2016,Tang2017}, double quantum dots\cite{Agarwalla2015,Yu2016}, or qubits\cite{Wang2017}. In these works, the generating function for the heat full-counting statistics has been determined\cite{Kindermann2004,Saito2007,Clerk2011,Agarwalla2012,Utsumi2014,Silaev2014,Berg2015,Agarwalla2015,Yu2016,Tang2017,Wang2017,Friedman2018,Ridley2019}, and the zero-frequency heat noise has been calculated\cite{Callen1951,Blencowe1999,Krive2001,Kindermann2004,Saito2007,Wu2009,Sanchez2013,Battista2014,Moskalets2014,Ridley2019} as well as the symmetrized finite-frequency heat noise\cite{Averin2010,Sergi2011,Zhan2011,Zhan2013,Pekola2018}. The non-symmetrized finite-frequency heat noise has also been calculated for a quantum dot (QD) but only for symmetrical couplings between the QD and the electron reservoirs\cite{Eymeoud2016}. The objective of the present work is twofold: first to generalize the calculation of non-symmetrized finite-frequency heat noise to the case of asymmetrical couplings, which can differ by a factor of up to ten\cite{Delagrange2018}, by looking at both auto-correlators and cross-correlators, and second to highlight the main features of the heat noise spectrum in a QD. Only the electronic contribution to the heat noise is considered in this work.

The paper is organized as follows: the model and results are presented in Sec.~II, the equilibrium and out-of-equilibrium heat noises are, respectively, discussed in Secs.~III and IV, and the conclusion is given in Sec.~V.


\section{Model and results} 

The standard Hamiltonian describing a non-interacting QD connected to left~(L) and right~(R) reservoirs of electrons is the Anderson Hamiltonian,
\begin{eqnarray}
 \mathcal{H}&=&\sum_{\alpha=L,R}\sum_{k\in \alpha}\varepsilon_{\alpha k} c^\dag_{\alpha k} c_{\alpha k}+\varepsilon_0 d^\dag d\nonumber\\
 &&+\sum_{\alpha=L,R}\sum_{k\in \alpha} (V_{\alpha k}c^\dag_{\alpha k} d +h.c.)
\end{eqnarray}
where $c^\dag_{\alpha k}$ ($d^\dag$), $c_{\alpha k}$ ($d$) are the creation and annihilation operators associated with the reservoir $\alpha$ (respectively the QD). The energies $\varepsilon_{\alpha k}$, $\varepsilon_0$ and $V_{\alpha k}$ are respectively the energy of the electrons in the reservoir $\alpha$, the discrete energy level of the QD, and the hopping integral between the reservoirs and the QD. The retarded Green function associated with the QD connected to the reservoirs is given in the flat wide-band limit by $G^r(\varepsilon)=(\varepsilon-\varepsilon_0+i(\Gamma_L+\Gamma_R)/2)^{-1}$, where $\Gamma_\alpha=2\pi \rho_\alpha|V_{\alpha}|^2 $ is the coupling between the QD and the reservoir $\alpha$ assuming that the density of states $\rho_\alpha$ and $V_{\alpha}\equiv V_{\alpha k}$ are energy independent.


\begin{widetext}

\begin{table}[b]
\renewcommand\arraystretch{1.5}
\begin{center}
\begin{tabular}{|l||c|c|c|c|}
\hline
$M_{\alpha\beta}^{\gamma\delta}(\varepsilon,\omega)$& $\gamma=\delta=L$& $\gamma=\delta=R$&$\gamma=L$, $\delta=R$&$\gamma=R$, $\delta=L$\\ \hline\hline
$\alpha=L$& $ \big|\mathcal{E}_{L}(\varepsilon- \omega)t_{LL}(\varepsilon) $& $\mathcal{E}^2_L(\varepsilon-\frac{ \omega}{2})$ & $ \big|\mathcal{E}_{L}(\varepsilon)$ & $ \big|\mathcal{E}_{L}(\varepsilon- \omega)$\\
$\beta=L$&$+\mathcal{E}_{L}(\varepsilon)t_{LL}^*(\varepsilon- \omega)$&$\times\mathcal{T}_{LR}(\varepsilon)\mathcal{T}_{LR}(\varepsilon- \omega)$&$-\mathcal{E}_L(\varepsilon-\frac{ \omega}{2})t_{LL}(\varepsilon)  \big|^2$&$-\mathcal{E}_L(\varepsilon-\frac{ \omega}{2})t_{LL}(\varepsilon- \omega)  \big|^2$\\
&$ -\mathcal{E}_L(\varepsilon-\frac{ \omega}{2})t_{LL}(\varepsilon)t_{LL}^*(\varepsilon- \omega)  \big|^2$&&$\times \mathcal{T}_{LR}(\varepsilon- \omega)$&$\times \mathcal{T}_{LR}(\varepsilon)$\\
 \hline
$\alpha=R$& $\mathcal{E}^2_R(\varepsilon-\frac{ \omega}{2})$ &$ \big|\mathcal{E}_{R}(\varepsilon- \omega)t_{RR}(\varepsilon) $ &  $ \big|\mathcal{E}_{R}(\varepsilon- \omega)$ & $ \big|\mathcal{E}_{R}(\varepsilon)$ \\
$\beta=R$&$\times\mathcal{T}_{LR}(\varepsilon)\mathcal{T}_{LR}(\varepsilon- \omega)$&$+\mathcal{E}_{R}(\varepsilon)t_{RR}^*(\varepsilon- \omega)$&$-\mathcal{E}_R(\varepsilon-\frac{ \omega}{2})t_{RR}(\varepsilon- \omega)  \big|^2$&$-\mathcal{E}_R(\varepsilon-\frac{ \omega}{2})t_{RR}(\varepsilon)  \big|^2$\\
&&$ -\mathcal{E}_R(\varepsilon-\frac{ \omega}{2})t_{RR}(\varepsilon)t_{RR}^*(\varepsilon- \omega)  \big|^2$&$\times\mathcal{T}_{LR}(\varepsilon)$&$\times \mathcal{T}_{LR}(\varepsilon- \omega)$\\
 \hline
 $\alpha=L$&$\mathcal{E}_R(\varepsilon-\frac{ \omega}{2})t_{LR}(\varepsilon)t_{LR}^*(\varepsilon- \omega) $&$\mathcal{E}_L(\varepsilon-\frac{ \omega}{2})t_{LR}^*(\varepsilon)t_{LR}(\varepsilon- \omega) $&$\Big[\mathcal{E}_L(\varepsilon)t_{LL}(\varepsilon)$&$\Big[\mathcal{E}_L(\varepsilon-\omega)t_{LL}^*(\varepsilon-\omega)$ \\
$\beta=R$&$\times\Big[\mathcal{E}_L(\varepsilon-\frac{ \omega}{2})t_{LL}^*(\varepsilon)t_{LL}(\varepsilon- \omega)$&$\times\Big[\mathcal{E}_R(\varepsilon-\frac{ \omega}{2})t_{RR}(\varepsilon)t_{RR}^*(\varepsilon- \omega)$&$-\mathcal{E}_L(\varepsilon-\frac{ \omega}{2})\mathcal{T}_{LL}(\varepsilon)\Big]$&$-\mathcal{E}_L(\varepsilon-\frac{ \omega}{2})\mathcal{T}_{LL}(\varepsilon-\omega)\Big]$\\
&$-\mathcal{E}_L(\varepsilon-\omega)t_{LL}^*(\varepsilon)$&$-\mathcal{E}_R(\varepsilon-\omega)t_{RR}(\varepsilon)$&$\times\Big[\mathcal{E}_R(\varepsilon-\omega)t_{RR}(\varepsilon-\omega)$&$\times\Big[\mathcal{E}_R(\varepsilon)t_{RR}^*(\varepsilon)$\\
&$-\mathcal{E}_L(\varepsilon)t_{LL}(\varepsilon- \omega)\Big]$&$-\mathcal{E}_R(\varepsilon)t_{RR}^*(\varepsilon- \omega)\Big]$&$-\mathcal{E}_R(\varepsilon-\frac{ \omega}{2})\mathcal{T}_{RR}(\varepsilon-\omega)\Big]$&$-\mathcal{E}_R(\varepsilon-\frac{ \omega}{2})\mathcal{T}_{RR}(\varepsilon)\Big]$\\
 \hline
 $\alpha=R$&$\mathcal{E}_R(\varepsilon-\frac{ \omega}{2})t_{LR}^*(\varepsilon)t_{LR}(\varepsilon- \omega) $&$\mathcal{E}_L(\varepsilon-\frac{ \omega}{2})t_{LR}(\varepsilon)t_{LR}^*(\varepsilon- \omega) $&$\Big[\mathcal{E}_L(\varepsilon)t_{LL}^*(\varepsilon)$&$\Big[\mathcal{E}_L(\varepsilon-\omega)t_{LL}(\varepsilon-\omega)$ \\
$\beta=L$&$\times\Big[\mathcal{E}_L(\varepsilon-\frac{ \omega}{2})t_{LL}(\varepsilon)t_{LL}^*(\varepsilon- \omega)$&$\times\Big[\mathcal{E}_R(\varepsilon-\frac{ \omega}{2})t_{RR}^*(\varepsilon)t_{RR}(\varepsilon- \omega)$&$-\mathcal{E}_L(\varepsilon-\frac{ \omega}{2})\mathcal{T}_{LL}(\varepsilon)\Big]$&$-\mathcal{E}_L(\varepsilon-\frac{ \omega}{2})\mathcal{T}_{LL}(\varepsilon-\omega)\Big]$\\
&$-\mathcal{E}_L(\varepsilon-\omega)t_{LL}(\varepsilon)$&$-\mathcal{E}_R(\varepsilon-\omega)t_{RR}^*(\varepsilon)$&$\times\Big[\mathcal{E}_R(\varepsilon-\omega)t_{RR}^*(\varepsilon-\omega)$&$\times\Big[\mathcal{E}_R(\varepsilon)t_{RR}(\varepsilon)$\\
&$-\mathcal{E}_L(\varepsilon)t_{LL}^*(\varepsilon- \omega)\Big]$&$-\mathcal{E}_R(\varepsilon)t_{RR}(\varepsilon- \omega)\Big]$&$-\mathcal{E}_R(\varepsilon-\frac{ \omega}{2})\mathcal{T}_{RR}(\varepsilon-\omega)\Big]$&$-\mathcal{E}_R(\varepsilon-\frac{ \omega}{2})\mathcal{T}_{RR}(\varepsilon)\Big]$\\
 \hline
\end{tabular}
\caption{Expressions of matrix elements $M_{\alpha\beta}^{\gamma\delta}(\varepsilon,\omega)$ appearing in the finite-frequency heat noise of Eq.~(\ref{heat_noise}), setting $\hbar=1$, where $t_{\alpha\beta}(\varepsilon)=i\sqrt{\Gamma_\alpha\Gamma_\beta}G^r(\varepsilon)$ is the transmission amplitude, $\mathcal{T}_{\alpha\beta}(\varepsilon)=|t_{\alpha\beta}(\varepsilon)|^2$ is the transmission coefficient, and $\mathcal{E}_\alpha(\varepsilon)=\varepsilon-\mu_\alpha$ is the difference between the energy $\varepsilon$ of the particle and the chemical potential in the reservoir $\alpha$.}
\label{table}
\end{center}
\end{table}

\end{widetext}

\begin{figure}[t]
\includegraphics[width=8cm]{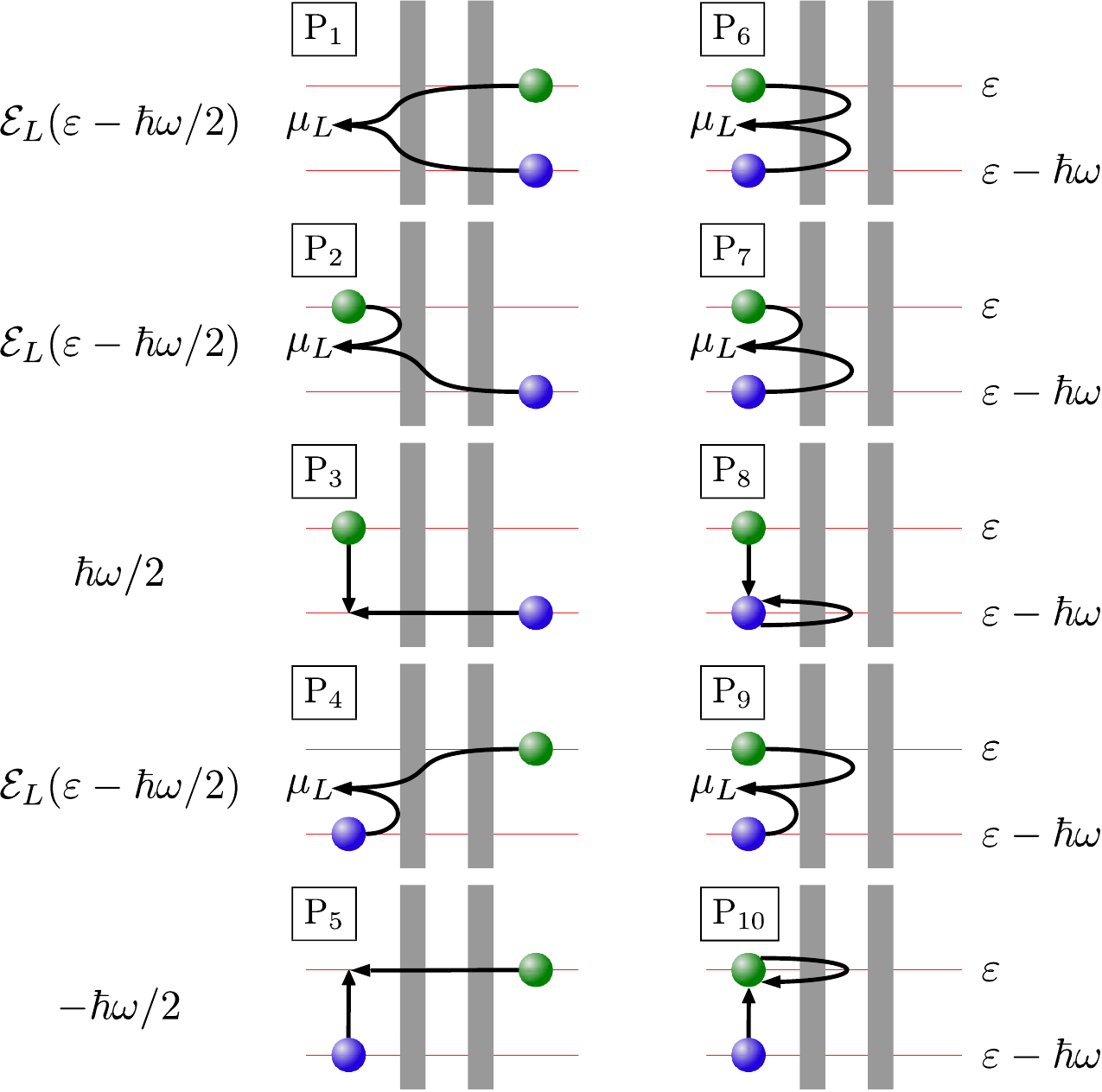}
\caption{Illustration of the 10 processes contributing to the heat noise in the left reservoir attached to the QD. The green (blue) spheres represent an electron (hole) of energy $\varepsilon$ ($\varepsilon-\hbar\omega$). The energy released for each process is indicated on the left side.}
\label{figure_processus}
\end{figure}

The heat noise is defined as the Fourier transform of the non-symmetrized correlator of heat currents at two different times: $\mathcal{S}^\text{heat}_{\alpha\beta}(\omega)=\int_{-\infty}^{\infty} \langle \Delta J_\alpha(t) \Delta J_\beta(0) \rangle e^{-i\omega t}dt$, where $\Delta J_\alpha(t)=J_\alpha(t)-\langle J_\alpha\rangle$. The heat current operator is given by\cite{note,Ludovico2014,Slimane2020,Ochoa2016} $J_\alpha(t)=-\mathcal{\dot H}_\alpha+\mu_\alpha \dot N_\alpha$, where $\mathcal{H}_\alpha=\sum_{k\in \alpha}\varepsilon_{\alpha k} c^\dag_{\alpha k} c_{\alpha k}$ is the Hamiltonian of the uncoupled reservoir $\alpha$ and $N_\alpha=\sum_{k\in \alpha}c^\dag_{\alpha k} c_{\alpha k}$ is the operator number of electrons in the reservoir $\alpha$. The calculation of the non-symmetrized finite-frequency heat noise is performed using the nonequilibrium Green function technique. It gives\cite{SM}
\begin{eqnarray}\label{heat_noise}
\mathcal{S}_{\alpha\beta}^\text{heat}(\omega)=\frac{1}{h}\sum_{\gamma, \delta=L,R}\int^{\infty}_{-\infty} d\varepsilon 
M^{\gamma\delta}_{\alpha\beta}(\varepsilon,\omega)f^e_\gamma(\varepsilon)f_\delta^h(\varepsilon-\hbar\omega)\nonumber\\
\end{eqnarray}
where $f_\gamma^e(\varepsilon)=(1+\exp((\varepsilon-\mu_\gamma)/k_BT_\gamma))^{-1}$ and $f_\delta^h(\varepsilon)=1-f_\delta^e(\varepsilon)$ are the Fermi-Dirac distributions for the electrons in the reservoir~$\gamma$ and the holes in the reservoir~$\delta$. $\mu_\gamma$ and $T_\gamma$ are  respectively the chemical potential and the temperature in the reservoir $\gamma$. The matrix elements $M^{\gamma\delta}_{\alpha\beta}(\varepsilon,\omega)$ are listed in Table \ref{table}. The relative importance of the four terms in Eq.~(\ref{heat_noise}) varies according to the experimental conditions: they all have an equal weight at equilibrium while the term $M^{LR}_{\alpha\beta}(\varepsilon,\omega)$ dominates out-of-equilibrium. The result of Eq.~(\ref{heat_noise}) is applicable to any frequency $\omega$, temperatures $T_{L,R}$, voltage $V$ and couplings $\Gamma_{L,R}$. It generalizes the results of Ref.~\onlinecite{Eymeoud2016} to arbitrary couplings between the QD and the reservoirs. The expressions of the elements for the matrix $M$ reduce to the ones entering in the expression of the charge noise $\mathcal{S}_{\alpha\beta}^\text{charge}(\omega)$ of Refs.~\onlinecite{Zamoum2016} and \onlinecite{Crepieux2018} provided that the factor $\mathcal{E}_\alpha(\varepsilon)$ is replaced by the value 1. One notices that three of such factors enter into the expression of the heat noise: $\mathcal{E}_\alpha(\varepsilon)$,  the energy of the electron in the reservoir~$\alpha$, $\mathcal{E}_\alpha(\varepsilon-\hbar\omega)$,  the energy of the hole in the reservoir~$\alpha$, and $\mathcal{E}_\alpha(\varepsilon-\hbar\omega/2)$, the average energy of the electron-hole pair in the reservoir~$\alpha$. These factors are related to the energy exchanged with the electromagnetic environment surrounding the QD during the various transfer processes contributing to the heat noise. These processes are 10 in number (for $\mathcal{S}_{LL}^\text{heat}(\omega)$, see Fig.~\ref{figure_processus}) and involve transfer of electron-hole pairs through the QD. Depending on the initial location of the electron and the hole, the number of possible processes differs. When the electron and the hole are both located in the right reservoir, there is only one process: P$_1$. When the electron is located in the left (right) reservoir and the hole in the right (left) reservoir, there are two processes: P$_2$ and P$_{3}$ (P$_4$ and P$_{5}$).  Finally, when the electron and the hole are both initially located in the left reservoir, there are five processes: from P$_{5}$ to P$_{10}$. To calculate the contribution of each of these sets of processes, one must take the quantum superposition of the processes having the same initial state\cite{SM}. Using this simple rule, one recognizes the expression of $\mathcal{S}_{LL}^\text{heat}(\omega)$ of Table~\ref{table}. It highlights the fact that the energy exchanged with the electromagnetic environment during the processes is the average energy of the electron-hole pair, i.e., either $\mathcal{E}_L(\varepsilon-\hbar\omega/2)$, when the electron and the hole both make an excursion in the central part of the QD, or $\pm \hbar\omega/2$, when only one of the two particles makes the excursion. This is the case for the processes P$_{3}$, P$_{5}$, P$_{8}$ and P$_{10}$, which have the particularity to contribute to the heat noise whereas they do not contribute to the charge noise\cite{Zamoum2016}.

\begin{figure}[t]
\includegraphics[width=8.7cm]{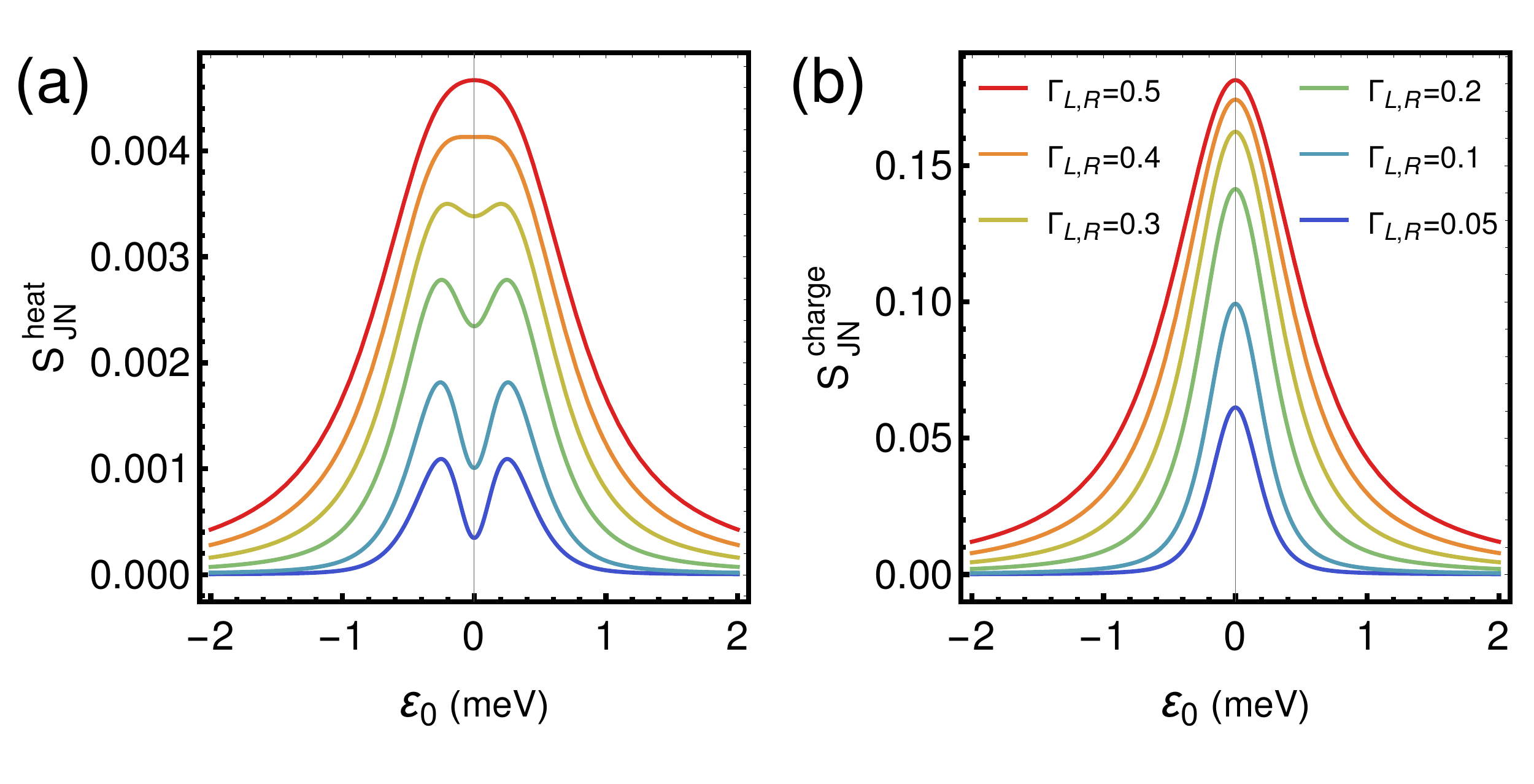}
\caption{Johnson-Nyquist (a) heat and (b) charge noises at equilibrium ($eV=\hbar\omega=0$ and $T_{L,R}=T$) as a function of the QD level energy $\varepsilon_0$ at $k_BT=0.1$~meV (i.e., $T=1.16$ K) for several values of the couplings between the reservoirs and the QD taken symmetrical: $\Gamma_{L,R}=\Gamma$ (in meV).
}
\label{figure0}
\end{figure}

\section{Equilibrium heat noise} 

 Before exploiting the result of Eq.~(\ref{heat_noise}), one checks that it gives the expected behavior for heat noise within known limits. At zero frequency $\omega=0$, symmetrical couplings $\Gamma_{L,R}=\Gamma$ with  $\mathcal{T}(\varepsilon)=\Gamma^2 G^r(\varepsilon)G^a(\varepsilon)$, and using the optical theorem that holds for a non-interacting QD, which means that one has $t(\varepsilon)+t^*(\varepsilon)=2\mathcal{T}(\varepsilon)$, Eq.~(\ref{heat_noise}) leads for the auto-correlators ($\alpha=\beta$) to the expression\cite{SM}
\begin{eqnarray}\label{heat_JN}
 &&\mathcal{S}_{\alpha\alpha}^\text{heat}(0)=\frac{1}{h}\int^{\infty}_{-\infty} d\varepsilon (\varepsilon-\mu_\alpha)^2\nonumber\\
 &&\times\bigg[\mathcal{T}(\varepsilon)(1-\mathcal{T}(\varepsilon))(f^e_\alpha(\varepsilon)-f^e_{\overline\alpha}(\varepsilon))^2\nonumber\\
 &&+\mathcal{T}(\varepsilon)(f^e_\alpha(\varepsilon)f^h_\alpha(\varepsilon)+f^e_{\overline\alpha}(\varepsilon)f^h_{\overline\alpha}(\varepsilon))\bigg]
\end{eqnarray}
in agreement with the results of Refs.~\cite{Krive2001,Battista2014,Tang2017}. The index ${\overline\alpha}$ takes the value $R$ for $\alpha=L$ and the value $L$ for $\alpha=R$. The last line of Eq.~(\ref{heat_JN}) corresponds to the equilibrium heat noise $ \mathcal{S}_\text{JN}^\text{heat}$ (Johnson-Nyquist), which can be expressed as a function of the thermal conductance  $K_\alpha=\partial\langle J_\alpha\rangle/\partial T_\alpha$ by the relation $\mathcal{S}_\text{JN}^\text{heat}=k_BT_L^2 K_L+k_B T_R^2 K_R$, in perfect agreement with Refs.~\cite{Krive2001,Saito2007,Crepieux2015}. One is reminded that the equilibrium charge noise $ \mathcal{S}_\text{JN}^\text{charge}$ is related to the electrical conductance by the relation $ \mathcal{S}_\text{JN}^\text{charge}=k_BT_L G_L+k_B T_R G_R$, with $G_\alpha=e\partial\langle I_\alpha\rangle/\partial\mu_\alpha$, where $\langle I_\alpha\rangle$ is the electrical current associated to the reservoir $\alpha$. The Johnson-Nyquist heat and charge noises are displayed in Figs.~\ref{figure0}(a) and (b) as a function of the QD energy level~$\varepsilon_0$. Within a certain range of values for the coupling $\Gamma$, $\mathcal{S}_\text{JN}^\text{heat}$ shows a double-peak profile while a single one is observed in $\mathcal{S}_\text{JN}^\text{charge}$. Indeed, at equilibrium, the charge fluctuations are maximal when the QD energy level is aligned with the chemical potentials, i.e. at $\varepsilon_0=0$ when $\mu_{L,R}=0$, since charge transfer does not cost energy. It results in a local minimum in the heat noise at $\varepsilon_0=0$. For increasing values of $|\varepsilon_0|$, the heat noise starts to increase because the charge transfer costs energy in this case. Then it finally decreases and converges to zero due to the fact that the probability for the charge to be transferred through the dot vanishes at high $|\varepsilon_0|$. When the two peaks in $\mathcal{S}_\text{JN}^\text{heat}$ are present, their positions are $\varepsilon_0\approx \pm 2.5 k_BT$ at most\cite{SM} in line with Ref.~\onlinecite{Tsaousidou2010} where such a double peak structure has been predicted in the thermal conductance of a QD. The condition to have a double peak in $ \mathcal{S}_\text{JN}^\text{heat}$ can be identified\cite{SM}. One finds that the condition is $\Gamma_L+\Gamma_R\lesssim 8k_BT$. These results could be verified experimentally since at equilibrium the heat noise is proportional to the thermal conductance.


\section{Out-of-equilibrium heat noise}

\begin{figure}[t]
\includegraphics[width=8.7cm]{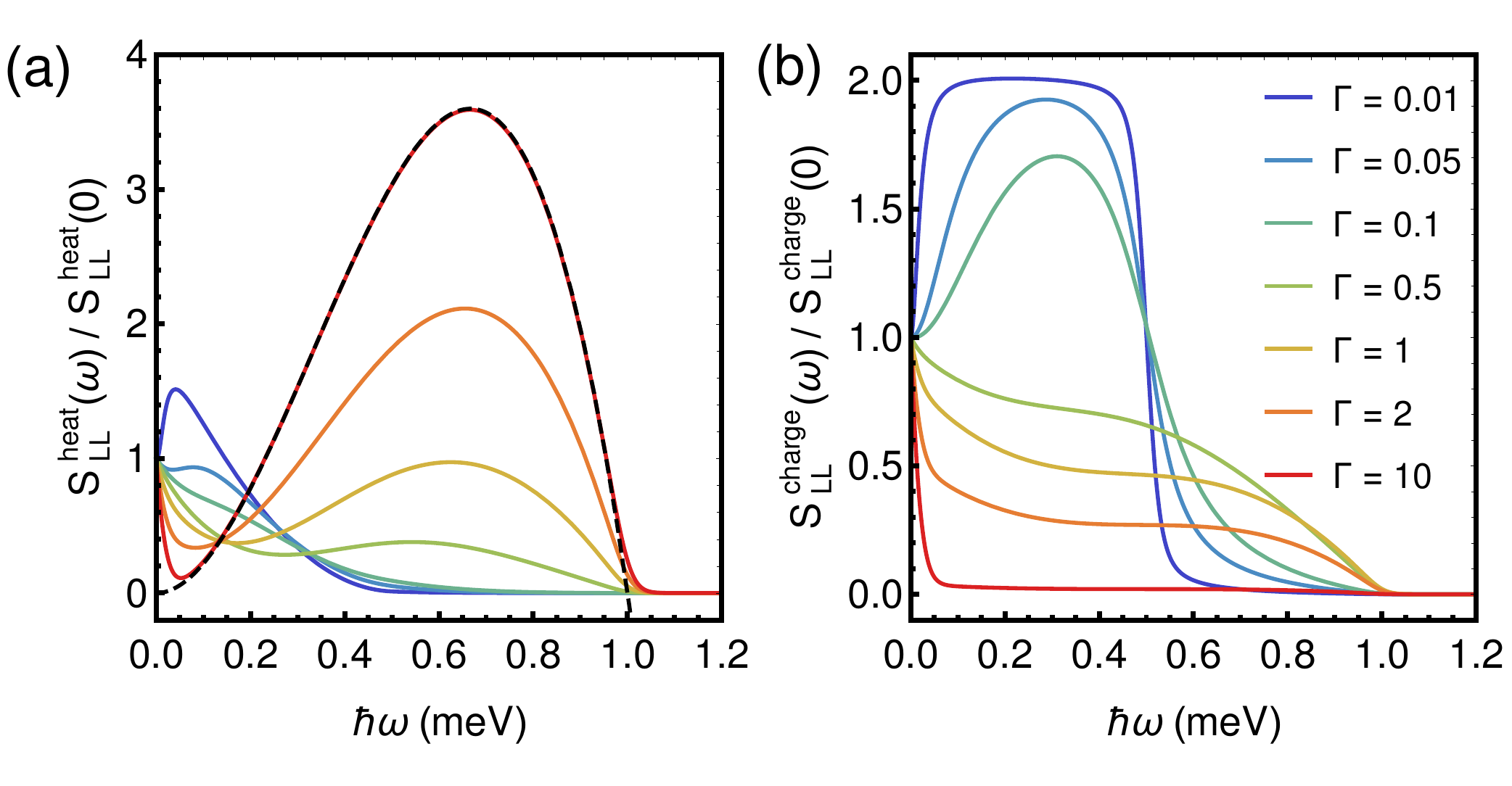}
\caption{Out-of-equilibrium (a)~heat noise and (b)~charge noise in the left reservoir as a function of frequency at $k_BT_{L,R}=0.01$ meV, $\varepsilon_0=0$, and $eV=1$ meV for different coupling values taken symmetrical:~$\Gamma=\Gamma_{L,R}$ (in meV). The dashed black line in (a) corresponds to Eq.~(\ref{heat_noise_high_transmission}).}
\label{figure_strong_coupling}
\end{figure}

The heat noise is sensitive to the fact that the system is driven out-of-equilibrium either by applying a bias voltage, a temperature gradient, or by considering the noise at finite frequency.  Figure~\ref{figure_strong_coupling} shows the profiles of heat and charge noises in the left reservoir as a function of the frequency at low temperature and fixed value of the voltage. For increasing values of the couplings $\Gamma_{L,R}$ (here taken symmetrical), the profile of the charge noise displayed in Fig.~\ref{figure_strong_coupling}(b) changes until it vanishes for strong couplings since the transmission becomes perfect in this limit ($\mathcal{T}_{LR}(\varepsilon)\approx 1$) which means that no fluctuation of charge current can occur: $ \mathcal{S}_{L L}^{\text{charge}}(\omega)=0$. This is not the case for heat noise. On the contrary, one observes that the heat noise globally increases when the couplings increase until it reaches the curve of equation
\begin{eqnarray}\label{heat_noise_high_transmission}
  \mathcal{S}_{\alpha\alpha}^{\text{heat}}(\omega>0,T=0)= \frac{\hbar^2\omega^2}{4h}(eV-\hbar\omega)\Theta(eV-\hbar\omega)
\end{eqnarray}
derived in Ref.~\onlinecite{SM} in the zero temperature and perfect transmission limits, corresponding to the dashed black line in Fig.~\ref{figure_strong_coupling}(a). Here, $\Theta$ is the Heaviside function. More generally the heat noise at perfect transmission is expressed as follows
\begin{eqnarray}
  \mathcal{S}_{\alpha\alpha}^{\text{heat}}(\omega)&=&\frac{\hbar\omega}{h}\left(\frac{\hbar^2\omega^2}{6}+e^2V^2+2e^2\mathcal{L}T^2\right)N(\hbar\omega)\nonumber\\
&&  +\frac{\hbar^2\omega^2}{4h}\sum_\pm(\hbar\omega\pm eV)N(\hbar\omega\pm eV)
\end{eqnarray}
with $N(\hbar\omega)$ the Bose-Einstein distribution function, and $\mathcal{L}=\pi^2k_B^2/3e^2$ the Lorenz number. This is a key result that generalizes the Planck's law for driven out-of-equilibrium quantum systems\cite{Wurfel1982,Dawson1984,Fedorovich2000,Gabelli2018,Martin2020}. It gives $\mathcal{S}_{\alpha\alpha}^{\text{heat}}(\omega)\propto\hbar^3\omega^3N(\hbar\omega)$ at zero voltage and low temperature, and Eq.~(\ref{heat_noise_high_transmission}) at positive voltage and zero temperature since one has $N_{T=0}(\varepsilon)=-\Theta(-\varepsilon)$. It means that the heat noise could be interpreted as the radiative power spectrum associated with the~QD, opening the route to its measure. Thus, contrary to the charge noise, the heat noise does not vanish within the perfect transmission limit. It is related to the fact that external sources of energy are supplied to the system, by the applied voltage or the frequency of the measurement device, resulting in fluctuations of the heat current, except when $\hbar\omega>eV$ since the system cannot deliver energy at a frequency higher than the voltage in the low-temperature limit\cite{Basset2010}.

\begin{figure}[t]
\includegraphics[width=8.7cm]{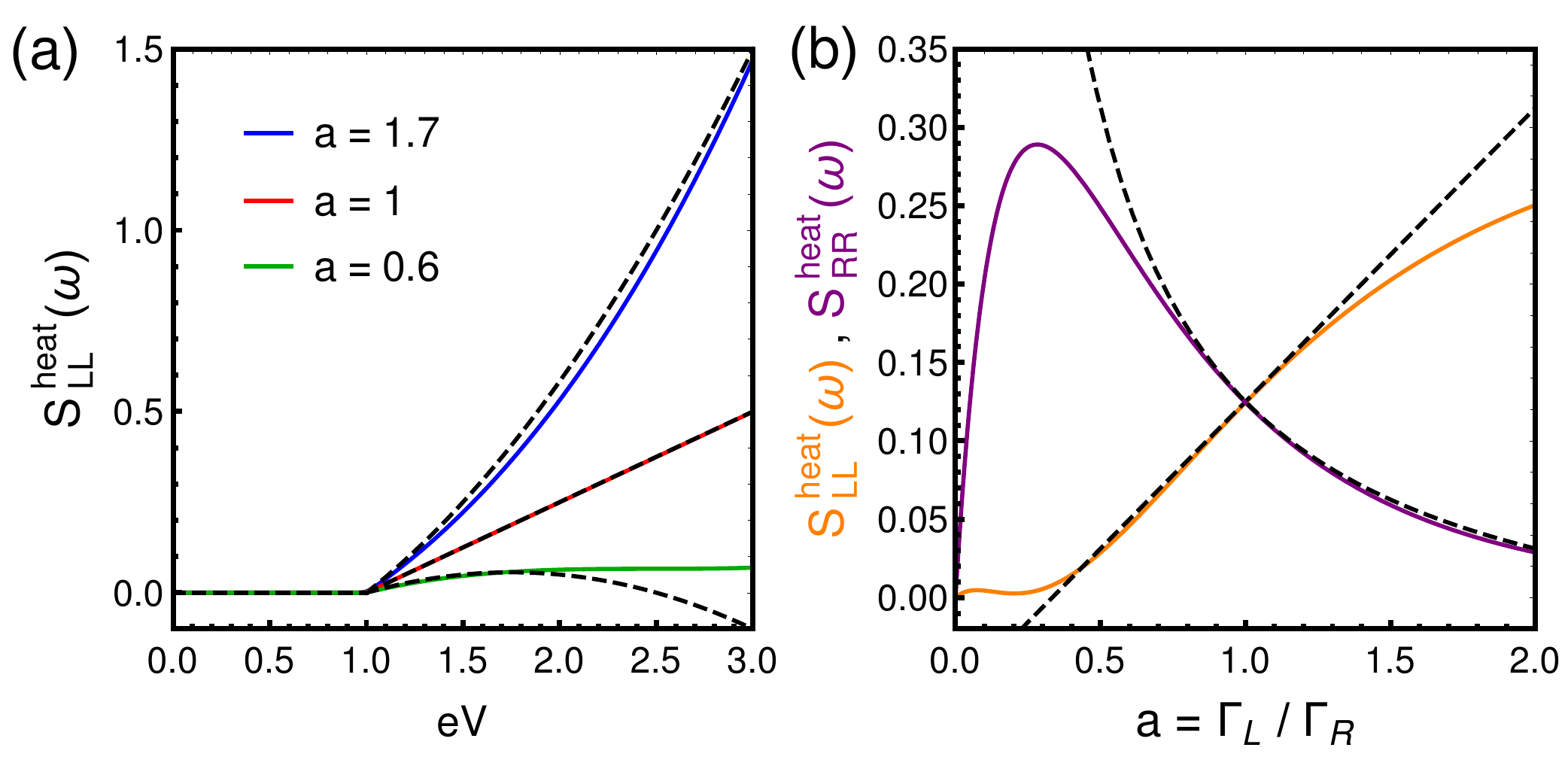}
\caption{Out-of-equilibrium (a) heat noise in the left reservoir as a function of voltage for three different values of the asymmetry factor $a=\Gamma_L/\Gamma_R$ with $\Gamma_L+\Gamma_R=40$~meV. (b)~Heat noise in the left and right reservoirs as a function of the asymmetry factor at $V=1.5$~mV and $\Gamma_R=10$ meV. The other parameters are $\hbar\omega=1$~meV, $k_BT_{L,R}=0.01$ meV, and $\varepsilon_0=0$. The dashed lines correspond to Eq.~(\ref{heat_noise_asym}).}
\label{figure_asymmetry}
\end{figure}

An additional distinctive feature between heat and charge noises is obtained for asymmetrical couplings $\Gamma_L\ne\Gamma_R$. Figure~\ref{figure_asymmetry}(a) shows the heat noise in the left reservoir as a function of voltage for different values of the asymmetry factor $a=\Gamma_L/\Gamma_R$ at low temperature and strong couplings. One observes that the variation of $\mathcal{S}_{L L}^{\text{heat}}(\omega)$ changes from a linear variation with $eV$ at $a=1$ to a quadratic variation with $eV$ at $a\ne 1$. Therefore, the fact that heat noise varies linearly or quadratically with voltage could provide information on the asymmetry of couplings. Note that this asymmetry only manifests itself when the QD is in a nonequilibrium situation when $eV$ and $\hbar\omega$ are both non-zero with the constraint $\hbar\omega < eV$. Such a linear/quadratic variation fits with the analytical expression, displayed by the dashed lines in Fig.~\ref{figure_asymmetry}(a), obtained for heat noise at zero temperature and perfect transmission\cite{SM}
\begin{eqnarray}\label{heat_noise_asym}
 \mathcal{S}_{\alpha\alpha}^{\text{heat}}(\omega>0,T=0)&=&\frac{\hbar\omega}{4h} (eV-\hbar\omega)\Theta(eV-\hbar\omega)\nonumber\\
 &&\times\left[\left(\frac{\Gamma_\alpha}{\Gamma_{\overline{\alpha}}}-1\right)eV+\hbar\omega\right]
\end{eqnarray}
This result can be written alternatively under the form $\mathcal{S}_{\alpha\alpha}^{\text{heat}}(\omega)=\Delta U  \Delta E/P_\alpha$, where $\Delta U=\hbar\omega/4$ is the average energy deviation in the equivalent RLC circuit including the electromagnetic environment\cite{Grabert1992}, $\Delta E=eV-\hbar\omega$ is the energy barrier that the charge has to overcome during the transfer processes, and $P_\alpha=(\omega/2\pi+(\Gamma_\alpha/\Gamma_{\overline{\alpha}}-1)eV/h)^{-1}$ is a period that characterizes the dynamics of the energy exchange associated to the reservoir $\alpha$: the dynamics is faster and the heat noise higher (see Fig.~\ref{figure_asymmetry}(b)) in the strongest connected reservoir since one has $P_\alpha<P_{\overline{\alpha}}$ when $\Gamma_\alpha>\Gamma_{\overline{\alpha}}$. For symmetrical barriers, i.e., $a=1$, the period is identical in the two reservoirs and is equal to $2\pi/\omega$, and the left and right heat noises coincide.

Experimentally, the heat noise at finite frequency could be obtained either from the measurement of the radiative power spectrum, as explained previously, or from the measurement of temperature fluctuations since one has the following relation\cite{Dashti2018,Karimi2020}:
$\mathcal{S}_{\alpha\alpha}^{\text{heat}}(\omega)=(1+\omega^2\tau_{E,\alpha}^2)K_\alpha^2\mathcal{S}_{\alpha\alpha}^{\text{temp}}(\omega)$,
where $ \mathcal{S}_{\alpha\alpha}^{\text{temp}}(\omega)=\int_{-\infty}^\infty \langle \Delta T_\alpha(t)\Delta T_\alpha(0)\rangle e^{-i\omega t}dt$. The energy relaxation time $\tau_{E,\alpha}$ is defined as the ratio between the heat capacity $C_\alpha=\partial \langle Q_\alpha\rangle/\partial T_\alpha$ and the thermal conductance $K_\alpha=\partial \langle J_\alpha\rangle/\partial T_\alpha$: $\tau_{E,\alpha}=C_\alpha/K_\alpha$. When only electrons contribute to the heat current, one has $\tau_{E,\alpha}=\hbar\Gamma_\alpha/(\Gamma_L\Gamma_R)$.


\section{Conclusion}

The study of electronic heat noise reveals several features that are not visible in charge noise. At equilibrium and provided that $\sum_\alpha\Gamma_\alpha \lesssim 8 k_BT$, the Johnson-Nyquist heat noise represented as a function of the QD energy level $\varepsilon_0$ shows a double-peak structure  instead of the single-peak structure visible in the charge noise. Out-of-equilibrium, since the QD can exchange energy with its electromagnetic environment, the heat noise does not vanish for perfect transmission, while the charge noise does, resulting in a crucial difference between these two quantities. Moreover, an out-of-equilibrium Planck's law is derived in that limit, meaning that the heat noise could be interpreted as the radiative power spectrum. Finally, unlike charge noise, heat noise is very sensitive to the coupling asymmetry, with a transition from a quadratic to a linear voltage variation when the couplings change from asymmetrical to symmetrical. A direct extension of this work would be the determination of the heat noise in an interacting QD using, for instance, the theory developed in Ref.~\onlinecite{Crepieux2018} for the calculation of charge noise. 


{\it Acknowledgments} -- The author would like to acknowledge T.Q.~Duong, P.~Eym\'eoud, G.~Fleury, J.~Gabelli, M.~Lavagna, F.~Michelini, M.~Moskalets, S.~Sahoo and R.~Zamoum for useful discussions.


\end{document}